# An experiment on the shifts of reflected C-lines

by W. Löffler[1], J. F. Nye[2] and J. H. Hannay[2]

[1] Huygens Laboratory, Leiden University, P.O. Box 9504, 2300 RA Leiden, The Netherlands

[2] H H Wills Physics laboratory, University of Bristol, Tyndall Avenue, Bristol BS8 1TL, U.K.

email: loeffler@physics.leidenuniv.nl

Abstract

An experiment is described that tests theoretical predictions on how C-lines incident obliquely on a surface behave on reflection. C-lines in a polarised wave are the analogues of the optical vortices carried by a complex scalar wave, which is the usual model for describing light and other electromagnetic waves. The centre of a laser beam that carries a (degenerate) C-line is shifted on reflection by the well-known Goos-Hänchen and Imbert-Fedorov effects, but the C-line itself splits into two, both of which are shifted longitudinally and laterally; their shifts are different from that of the beam centre. To maximise the effect to be measured, internal reflection in a glass prism close to the critical angle was used. In a simple situation like this two recently published independent theories of C-line reflection overlap and it is shown that their predictions are identical. The measured differences in the lateral shifts of the two reflected C-lines are compared with theoretical expectations over a range of incidence angles.

1. **Introduction**

Wave dislocations [1, 2] or optical vortices as they are now called, lines of zero disturbance, are features in the complex scalar fields commonly used to represent light. The phase circulates around them in the manner of a vortex. They occur generically, that is, they appear naturally in any general monochromatic scalar field. However, in the vector electromagnetic field that fully describes light, including its polarization, the end of the electric vector at each point describes a polarization ellipse, which collapses to zero only at special points; zeros are not generic. A feature in a vector wavefield that is generic and may be thought of as the counterpart of the line vortex in scalar waves is a line where the polarization ellipse is a circle, a $C_T$-line (T for true). The polarization ellipse observed on a screen is not this but its projection. Points appear on the screen, C-points, where the observed projected polarization is a circle, and as the screen is moved parallel to itself the points trace out lines, C-lines, in space. It is these C-lines that are the concern of this paper.

Previous papers [3-5] examined theoretically what happens when a laser beam containing a C-line is incident at an angle on a single plane interface between two different media. The usual Fresnel coefficients cannot be used directly here because they apply to a simple plane wave. Instead, the first of these theoretical papers and the last pair describe alternative schemes, each being a slightly different idealization of the core of the beam and each having its own advantages. The two schemes overlap, however, for a suitable simple beam, and this case is considered here – the experiment thereby acts to confirm both. It suffices to describe this overlap case by one of the two schemes—we choose the differentiated plane wave (DPW) approach and relate it to the other scheme briefly in an Appendix.

We may start with the monochromatic scalar plane wave $\psi = \exp(i\mathbf{k}\cdot\mathbf{r})$ and differentiate with respect to the direction of the wave, specified by the components of the wavevector $\mathbf{k} = (k_x, k_y, k_z)$, rather than with respect to position, given by the vector $\mathbf{r} = (x, y, z)$. After the differentiation we specialize to the $z$ direction, as follows:

$$\left(-i\frac{\partial \psi}{\partial k_x} + \frac{\partial \psi}{\partial k_y}\right)_{\mathbf{k}=(0,0,k_z)} = (x+iy)\exp[ik_z z] \quad (1)$$

which is an optical vortex, or a pure screw wave dislocation. The differentiation with respect to direction has produced from a featureless plane wave a wave with a linear modulation of amplitude across its wavefronts; it is zero on $x = y = 0$, which is the dislocation or vortex line. A modulation of amplitude would normally lead to diffraction but with this linear modulation the wave propagates unchanged in form. Notice that the original plane wave obeys the Helmholtz equation $\nabla^2\psi + k^2\psi = 0$ and so the differentiated plane wave automatically does so too. A similar, but necessarily not so simple, treatment applies to vector electromagnetic waves; the original plane wave obeys Maxwell's equations and the differentiated plane wave automatically does so too. (For the electric field this means not only satisfying the wave equation for each component separately but also the condition div $\mathbf{E}$ =0. This feature makes the procedure very convenient in dealing with problems of transmission and reflection, for one is then assured of satisfying Maxwell's equations.)

The theory that followed from this approach made predictions about the shifts of C-lines that are analogous to, but not the same as, the Goos-Hänchen [6] and Imbert-Fedorov [7] shifts of the centroids of narrow laser beams. It was then natural to consider an experiment to test these theoretical predictions.

In the theory the incident wave containing the C-lines is infinitely wide. The laser beam used in the experiment had a Gaussian envelope that was comparatively narrow, but it proved to be wide enough for practical purposes. There was a more important difference. In the theory, which describes the fields actually present in space, the incident wave interferes with the reflected wave to produce an interference pattern in front of the interface, but an experiment set up to observe the reflected wave will not see this because the incident wave is travelling in a quite different direction. What it will see is just the reflected wave in front of the interface and its virtual field behind it. The theory was modified to take this difference into account.

The shifts are often only a few wavelengths. Various ways of producing an effect large enough to be measured accurately were considered. For example, a Fabry-Perot type slab where the multiple reflections can enhance the effect. But this required a measurement of the thickness accurate to a fraction of a wavelength and so was not readily practicable. Finally, it was decided to work with total internal reflection close to the critical angle and this does give a large enough effect. It should be noted that the theory calculates the spatial shifts that occur at the interface itself and makes no prediction about any purely angular deviation, whereas the experiment measures the shift at a small distance from the surface. The comparison assumes, it seems plausibly, that any purely angular deviation may be neglected. Section 2 describes the experiment in detail.

## 2. Experiment

The optical beams used in a realistic experiment are characterized by a plane-wave superposition with finite spread around a main beam propagation direction, $z'$. They can be adequately described in the paraxial approximation, where the transverse beam amplitude changes little along the propagation direction, except for the oscillating phase $\exp(ikz')$ ($k = \omega/c$ is the wave vector and ω the angular frequency). A differentiated plane wave describing the core of a simple vortex Laguerre-Gauss beam can be written [4, 5] in the beam-local coordinate frame $x', y', z'$ as

$$\left.\begin{aligned} E'_x &= k(x' + iy')\exp[ikz'] \\ E'_y &= ik(x' + iy')\exp[ikz'] \\ E'_z &= 0 \end{aligned}\right\} \quad (2)$$

This (degenerate) C-line in an electromagnetic wave corresponds to a vortex $r'\exp(i\phi')$ in a scalar wave, where $r', \phi'$ are polar coordinates in the local beam frame. The full Laguerre-Gauss beam is the differentiated plane wave of equation (2) multiplied by a Gaussian envelope $\exp(-r^2/w^2)$ of waist $w$. We make this beam in the laboratory simply by creating a $\ell=1$ orbital angular momentum (OAM) [8] beam with homogeneous circular polarization. As shown in Fig. 1, we use a reflective phase-only spatial light modulator to create the optical vortex, and a quarter-wave plate to transform linear to circular polarization.

The beam is internally reflected at the hypotenuse face of a BK7 ($n = 1.5106$) prism, where correction of the angle of incidence for refraction at the prism legs is done. After reflection, we use a combination of quarter-wave plate, linear polarizer, and CCD camera (pixel size 4.4 μm) to analyze the reflected field for left and right circular polarization components. We find C-lines in both beams, and to determine their position with high accuracy, we use computer analysis: we fit a fourth-order polynomial to the field in the vicinity of the vortex and determine its centre position. Since we have no reference, we determine the differential position of the two C-lines by switching the analysing wave plate WP2 between ±45°. The accuracy of this method is around 10 μm, which we have estimated from repeated measurements.

Fig. 2 shows the measured C-line shift as a function of the angle of incidence, compared with theoretical prediction. We see that, as the incident angle passes through the critical angle, the shift diverges and changes sign—as theory predicts. The apparent systematic deviation is probably due to interface imperfections, as has been observed many times before (e.g., [9, 10]).

## 3. Discussion

We have demonstrated experimentally how C-lines, points of pure apparent circular polarization, are displaced upon reflection at planar interfaces.

It is striking to see that our experiment with realistic, finite optical beams agrees very well with theory, despite the fact that the theory is based on infinitely extended fields, i.e., differentiated plane waves. This confirms that only the field in the vicinity of the optical vortex determines the C-line shift, and that the Gaussian envelope is less relevant. We note that, in our experiment, we did not have to use a very sensitive position detection method such as a split detector in combination with a lock-in amplifier as is common in beam-shift experiments [11], because the vortex core position can be determined with very high accuracy. More than 60 years ago Wolter [12], suggested that the nodal line of a pair of plane waves [13] should be easier to detect than the centroid of an optical beam; our experimental results confirm that suggestion.

Our work therefore connects the field of polarization singularities [14], which is based on exact solutions of the 3D Maxwell equations, to optical beam shifts [15], when they are described within complex scalar optics; in particular to shifts of beams with orbital angular momentum [9, 16, 17]. Beam shifts describe deviations from geometrical optics due to diffractive corrections, and our findings here demonstrate that observation of C-lines leads to similar results.

We have shown that our observed displacement of a vortex closely follows the prediction of both the theoretical schemes referred to in the introduction. As a final remark, the observed displacement of a fundamental vortex could alternatively be described by the appearance of "orbital angular momentum sidebands" [18], which appear if the OAM spectrum is determined with respect to the axis given by geometrical reflection. The observed C-line shift implies a broadened OAM spectrum.

Acknowledgements: We thank J.Götte and M.Dennis for fruitful discussions on the properties of C-lines.

**Appendix: The C-line shift formulas**

The two separate theoretical schemes, Dennis-Götte [3] and Hannay-Nye [4, 5], for describing the core of the beam that carries the C-line are both based on series (Taylor) expansion of the local electric field. Both have the property that although the polarization varies locally, it becomes uniform away from the locality, representing the main polarization of the beam.

The first scheme, which might be called the 'analytic polynomial wave' (APW) uses the fact that for Laguerre-Gauss beams the local field can be described by a polynomial function of the single complex variable $x+iy$ where $x,y$ are the transverse coordinates. Higher-order beams are described by higher-order polynomials. The local analyticity feature is preserved in reflection from an interface (but not in transmission). The second scheme, the differentiated plane wave (DPW) does not assume analyticity (but does not exclude it). A reflected DPW is still a DPW (as is a transmitted one although that does not arise here). The DPW only deals with the lowest non-trivial level of series expansion – the first order, appropriate to the simplest beam containing a C-line—but that is all that is required here.

The incident beam in the experiment is the lowest order Laguerre-Gauss beam (beyond a simple Gaussian), having a central zero line of electric field and uniform circular polarization everywhere. Its core can be described by either scheme (it is a highly degenerate C-line). The reflected beam as a whole is no longer a Laguerre-Gauss beam, but its core can again be described by either the DPW or the APW scheme, having a single straight right-handed C-line and a single straight left-handed one. Both are parallel to, but shifted away from, the hypothetical straight line that would correspond to specular reflection of the incident zero line vortex.

The paper describing the APW scheme mostly analyses not the vortices or C-lines, but the centroid of the beam intensity –historically this has been the main focus of shift analysis. A short section, however, supplies explicitly the formula for the shifts of the two C-lines in reflection of an APW beam. The formula for the C-line shifts in a DPW was described in brief in [4, 5] but not carried through, so an explicit general derivation is given now. Where an APW is appropriate the result for a DWP is the same as that for an APW.

The reflected wave is generated from the incident Fourier spectrum of plane waves, each modified by multiplication by its own Fresnel reflection coefficient depending on its direction of propagation or wavevector **k**. Let this reflected Fourier spectrum of plane waves be denoted **E**(**k**)exp(i**k.r**). Its complete form will not be required in our task of finding the local form around the central zero line. This local form is supplied by the differentiated plane wave for the reflected beam: $\Delta \mathbf{k} \cdot \frac{\partial}{\partial \mathbf{k}}[\mathbf{E}(\mathbf{k})\exp(i\mathbf{k}\cdot\mathbf{r})]$ evaluated at the central wavevector **k** of the reflected beam. The constant complex vector $\Delta \mathbf{k}$ is a vector perpendicular to **k** determined by the incident differentiated plane wave; both **k** and $\Delta \mathbf{k}$ are mirror reflections in the interface plane of $\mathbf{k}_{inc}$ and $\Delta \mathbf{k}_{inc}$.

We digress briefly to describe the incident wave, (2), more formally. It can be considered as arising from the differentiation of a bundle of plane waves with 'constant' polarization (circular in the case of (2)). Thus starting with a form $\Delta \mathbf{k}_{inc} \cdot \frac{\partial}{\partial \mathbf{k}_{inc}}[\mathbf{E}_{inc}(\mathbf{k}_{inc})\exp(i\mathbf{k}_{inc}\cdot\mathbf{r})]$ one requires $\Delta \mathbf{k}_{inc} \cdot \frac{\partial}{\partial \mathbf{k}_{inc}} \mathbf{E}_{inc}(\mathbf{k}_{inc}) = 0$ by choice, representing the 'constant' polarization condition of the bundle. After the remaining differentiation of the exponential, the *z'* axis is defined to be along **k**$_{inc}$ and one takes $\mathbf{E}_{inc} = \{1, i, 0\}$ and $\Delta \mathbf{k}_{inc} = \{1, i, 0\}$ in the {*x',y',z'*} frame. This gives the wave specified by (2).

The circular (projected) polarization desired (which may be the same or opposite to that of the incident wave) is selected by taking the dot product of the electric field **E**(k) with a filter polarizer vector which in the case we have considered is {1, ±i, 0}. However, for generality and notational consistency with [3], it will be denoted by **F**[*]. Taking the dot product of the vector expression above with **F**[*] gives the scalar field $\left[\mathbf{r} - i\partial\log(\mathbf{E}\cdot\mathbf{F}^*)/\partial\mathbf{k}\right]\cdot\Delta\mathbf{k} \; i(\mathbf{E}\cdot\mathbf{F}^*)\exp(i\mathbf{k}\cdot\mathbf{r})$

whose zero line $\mathbf{r} = 0$ we seek, namely the solution of $[\ ]\cdot \Delta\mathbf{k} = 0$. Define $\mathbf{R} \equiv \mathbf{R_{Re}} + i\mathbf{R_{Im}} \equiv \partial \log(\mathbf{E}\cdot\mathbf{F}^*)/\partial\mathbf{k}$ and $\Delta\mathbf{k} \equiv \Delta\mathbf{k_{Re}} + i\Delta\mathbf{k_{Im}}$. Then $[\mathbf{r} - i\mathbf{R_{Re}} + \mathbf{R_{Im}}]\cdot(\Delta\mathbf{k_{Re}} + i\Delta\mathbf{k_{Im}}) = 0$ whose real and imaginary equations are simultaneous linear equations for $\mathbf{r}\cdot\Delta\mathbf{k_{Re}}$ and $\mathbf{r}\cdot\Delta\mathbf{k_{Im}}$. Now if we add any complex multiple of $\mathbf{k}$ to $\mathbf{r}$, it makes no difference to the two equations, and it therefore follows that the solution for $\mathbf{r}$ describes a straight line parallel to the vector $\mathbf{k}$. After some reduction the solution can be expressed in terms of an outer product matrix,

$$(\mathbf{r} + \mathbf{R_{Im}}) \times \mathbf{k} = \frac{k^2}{(\Delta\mathbf{k_{Re}} \times \Delta\mathbf{k_{Im}})\cdot\mathbf{k}} \left[\Delta\mathbf{k_{Re}} \otimes \Delta\mathbf{k_{Im}} + \Delta\mathbf{k_{Im}} \otimes \Delta\mathbf{k_{Re}}\right] \mathbf{R_{Re}}. \quad (A1)$$

The left hand side does not give $\mathbf{r}$ directly but $\mathbf{r}$ rotated by the cross product with $\mathbf{k}$. This result for the C-line shift in the DPW scheme is general. It overlaps with the result from the APW scheme when it can be expressed in terms of a complex variable x+iy. This is so for the particular case, equation (2), considered here, where $\Delta\mathbf{k} = \{1, i, 0\}$ so that $\Delta\mathbf{k_{Re}} = \{1, 0, 0\}$ and $\Delta\mathbf{k_{Im}} = \{0, 1, 0\}$. Thus the vectors $\Delta\mathbf{k_{Re}}$ and $\Delta\mathbf{k_{Im}}$ are of equal magnitude and at right angles (and also at right angles to $\mathbf{k}$). These magnitude and direction relations do not need to hold for a general DWP, indeed they are just the conditions that guarantee the form $x + iy$ required for an APW, equation (2) being a particular example. It follows that on the right hand side of (A1) there is first a multiplying factor and in the square brackets a multiple of the identity matrix. The right hand side is then a multiple of $\mathbf{R_{Re}}$, in fact $k\mathbf{R_{Re}}$. We have reached our objective because this makes (A1) the same as the formula previously obtained from the APW scheme, (equation 4 in [3]), which is in a rather different notation. To relate the notations note that the cross product in our equation is represented by the right-angle rotation operator $i\sigma_2$. On this common ground the two formulas agree.

## References

[1] Nye, J. F. & Berry, M. V. Dislocations in Wave Trains. *Proc. R. Soc. Lond. A* 336, 165 (1974).

[2] Nye, J. F. Natural focusing and fine structure of light: caustics and wave dislocations. *CRC Press* (1999).

[3] Dennis, M. R. & Götte, J. B. Topological Aberration of Optical Vortex Beams: Determining Dielectric Interfaces by Optical Singularity Shifts. *Phys. Rev. Lett.* 109, 183903 (2012).

[4] Hannay, J. H. & Nye, J. F. Refraction of C-line vortices. *J. Opt.* 15, 014008 (2013).

[5] Hannay, J. H. & Nye, J. F. A differentiated plane wave: its passage through a slab. *J. Opt.* 15, 044025 (2013).

[6] Goos, F. & Hänchen, H. Ein neuer und fundamentaler Versuch zur Totalreflexion. *Ann. Phys.* 436, 333-346 (1947).

[7] Imbert, C. Calculation and Experimental Proof of the Transverse Shift Induced by Total Internal Reflection of a Circularly Polarized Light Beam. *Phys. Rev. D* 5, 787 (1972).

[8] Allen, L., Beijersbergen, M. W., Spreeuw, R. J. C., & Woerdman, J. P. Orbital angular momentum of light and the transformation of Laguerre-Gaussian laser modes. *Phys. Rev. A* 45, 8185 (1992).

[9] Merano, M., Hermosa, N., Woerdman, J. P., & Aiello, A. How orbital angular momentum affects beam shifts in optical reflection. *Phys. Rev. A* 82, 023817 (2010).

[10] Löffler, W., Hermosa, N., Aiello, A., & Woerdman, J. P. Total internal reflection of orbital angular momentum beams. *J. Opt.* 15, 014012 (2013).

[11] Gilles, H., Girard, S., & Hamel, J. Simple technique for measuring the Goos-Hänchen effect with polarization modulation and a position-sensitive detector. *Opt. Lett.* 27, 1421 (2002).

[12] Wolter, H. Concerning the path of light upon total reflection. *J. Opt. A* 11, 090401 (2009).

[13] Dennis, M. R. & Götte, J. B. Beam shifts for pairs of plane waves. *Journal of Optics* 15, 014015 (2013).

[14] Nye, J. F. Lines of Circular Polarization in Electromagnetic Wave Fields. *Proc. R. Soc. Lond. A* 389, 279 (1983).

[15] Bliokh, K. Y. & Aiello, A. *J. Opt.* 15, 014001 (2013).

[16] Götte, J. B. & Dennis, M. R. Generalized shifts and weak values for polarization components of reflected light beams. *New J. Phys.* 14, 073016 (2012).

[17] Aiello, A. Goos-Hänchen and Imbert-Fedorov shifts: a novel perspective. *New J. Phys.* 14, 013058 (2012).

[18] Löffler, W., Aiello, A., & Woerdman, J. P. Observation of Orbital Angular Momentum Sidebands due to Optical Reflection. *Phys. Rev. Lett.* 109, 113602 (2012).

# Figures

**Figure 1:**

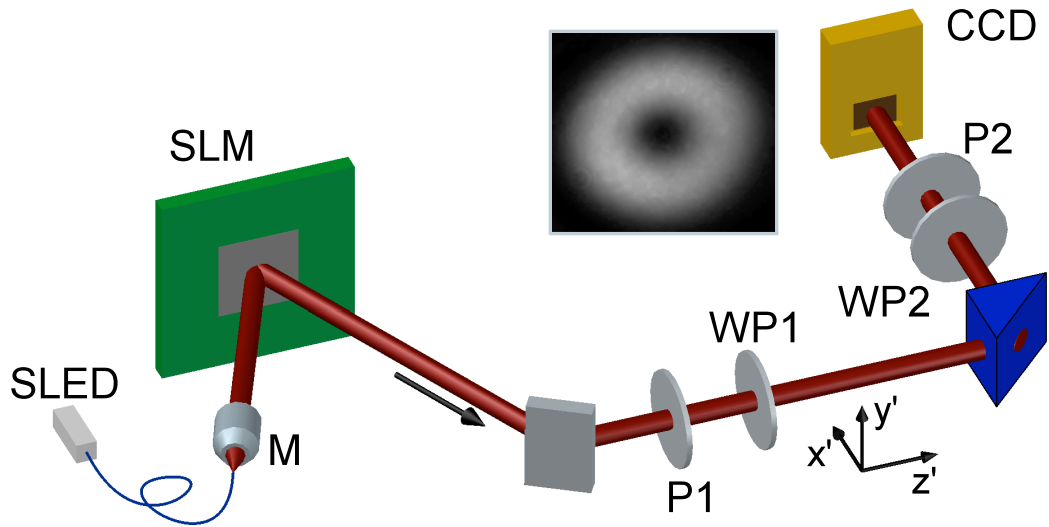

Fig. 1: Experimental setup. Light from a single mode fiber-coupled superluminescent diode (SLED, 825 ± 7 nm) is collimated using a 10x, NA 0.1 microscope objective (M). A topological phase of charge one is imprinted using a spatial light modulator (SLM); this vortex beam is then circularly polarized using horizontal polarizer P1 and quarter-wave plate WP1. After internal reflection at the BK7 prism (refractive index 1.5106), the beam is analyzed using a CCD camera after passing another quarter-wave plate WP2 and a linear polarizer P2. The inset shows a typical image of a C-line.

**Figure 2:**

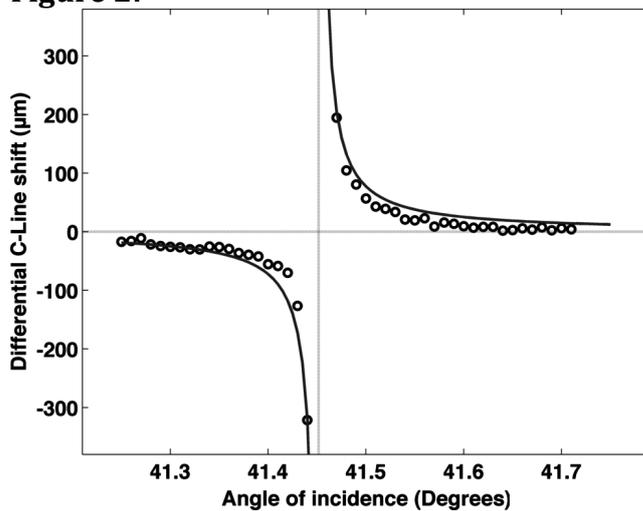

Fig. 2: Predicted (full line) and measured (circles) differential C-line shift around the critical angle for internal reflection. We see that the shift diverges and changes sign at critical incidence (vertical line).